\documentstyle[11pt,epsfig]{article}  
\def\simge{\mathrel{%
   \rlap{\raise 0.511ex \hbox{$>$}}{\lower 0.511ex \hbox{$\sim$}}}}   
\def\simle{\mathrel{   
   \rlap{\raise 0.511ex \hbox{$<$}}{\lower 0.511ex \hbox{$\sim$}}}}   
\def\slashchar#1{\setbox0=hbox{$#1$}           
   \dimen0=\wd0                                 
   \setbox1=\hbox{/} \dimen1=\wd1               
   \ifdim\dimen0>\dimen1                        
      \rlap{\hbox to \dimen0{\hfil/\hfil}}      
      #1                                        
   \else                                        
      \rlap{\hbox to \dimen1{\hfil$#1$\hfil}}   
      /                                         
   \fi}                                         %
   
\def\dfrac#1#2{{\displaystyle {#1 \over #2}}}

\def\simge{\mathrel{%
   \rlap{\raise 0.511ex \hbox{$>$}}{\lower 0.511ex \hbox{$\sim$}}}}   
\def\simle{\mathrel{   
   \rlap{\raise 0.511ex \hbox{$<$}}{\lower 0.511ex \hbox{$\sim$}}}}   
\def\slashchar#1{\setbox0=\hbox{$#1$}           
   \dimen0=\wd0                                 
   \setbox1=\hbox{/} \dimen1=\wd1               
   \ifdim\dimen0>\dimen1                        
      \rlap{\hbox to \dimen0{\hfil/\hfil}}      
      #1                                        
   \else                                        
      \rlap{\hbox to \dimen1{\hfil$#1$\hfil}}   
      /                                         
   \fi}                                         %
   
\def\dfrac#1#2{{\displaystyle {#1 \over #2}}}

\newcommand{\ifig}[1]{\mbox{\epsfig{file=#1,height=12cm,width=14cm}}}
\newcommand{\ifigsm}[1]{\mbox{\epsfig{file=#1,height=8cm,width=12cm}}}

\newcommand{\be}{\begin{equation}}   
\newcommand{\ee}{\end{equation}}   
\newcommand{\bea}{\begin{eqnarray}}   
\newcommand{\eea}{\end{eqnarray}}

\newcommand{\cals}{{\cal S}}   
\newcommand{\calo}{{\cal O}}

\newcommand{\RIMOM}{\mbox{\scriptsize RI/MOM}}

\newcommand{\Tr}{\mbox{Tr}\;}   
\newcommand{\Pj}{\mbox{I}\!\!\mbox{P}}   

\begin{document}   
\pagestyle{empty}    
\vspace{-0.6in}   
\begin{flushright}   
BUHEP-00-10\\
ROM2F/2000/21 \\
\end{flushright}   
\vskip 1.0in   
\centerline{\Large{\bf RI/MOM Renormalization Window}}
\centerline{\Large{\bf and Goldstone Pole Contamination}}   
\vskip 0.6cm   
\centerline{\bf{L.~Giusti$^{a}$ and A.~Vladikas$^{b}$}}   
\vskip 0.5cm   
\centerline{$^a$ Dept. of Physics, Boston University,}   
\centerline{590 Commonwealth Avenue, Boston MA 02215, USA.}   
\vskip 0.2cm
\centerline{$^b$ INFN, Sezione di Roma II,
c/o Dip. di Fisica, Univ. di Roma ``Tor Vergata'',}   
\centerline{Via della Ricerca Scientifica 1, I-00133 Roma, Italy.}

\vskip 1.5in
\begin{abstract}
We perform a comparative study of the ratio of lattice (Wilson fermion)
renormalization constants $Z_P/Z_S$, obtained non-perturbatively
from the RI/MOM renormalization
conditions and from Ward Identities of on- and off-shell Green's functions.
The off-shell Ward Identity used in this work relies on 
correlation functions with non-degenerate 
quark masses. We find that, due to discretization effects, 
there is a 10\%-15\% discrepancy between the two Ward Identity determinations
at current bare couplings ($\beta$ values). The RI/MOM result is in the same range
and has a similar systematic error of 10-15\%. Thus, contrary to a previous claim,
the contamination of the RI/MOM result from the presence of a Goldstone pole 
at renormalization scales $\mu \sim a^{-1}$ is subdominant, compared to finite cutoff effects.
\end{abstract}
\vfill
\pagestyle{empty}\clearpage   
\setcounter{page}{1}   
\pagestyle{plain}   
\newpage   
\pagestyle{plain} \setcounter{page}{1}   
   
\section{Introduction}   
\label{sec:intro}
The Non-Perturbative (NP) renormalization technique proposed in \cite{np}
has been successfully applied to compute renormalization constants 
of two-fermion \cite{ggrt1,sfqcd} and four-fermion \cite{z4f} operators
with Wilson fermions. The standard implementation consists in imposing 
the RI/MOM renormalization conditions on conveniently defined 
amputated projected Green functions computed between off-shell quark states of
operators of interest, evaluated numerically at fixed momentum and 
lattice spacing (fixed lattice coupling $\beta$). In order to control discretization effects, 
one should work at renormalization scales $\mu$ well below the UV cutoff 
(i.e. $\mu \ll \calo(a^{-1})$). 
Moreover, any contribution in the renormalization constants due to 
spontaneous chiral symmetry breaking terms in the Green's functions has to 
be avoided to ensure that the non-perturbatively renormalized operators 
satisfy the standard QCD axial Ward Identities  (abbreviated as WIs), 
i.e.  $\Lambda_{QCD} \ll \mu$. Thus, the existence of a 
``renormalization window''
$\Lambda_{QCD} \ll \mu \ll \calo(a^{-1})$ is essential to the reliability
of the RI/MOM scheme in practical simulations, as was pointed out in
ref.~\cite{np}. We note in passing that, as discussed in refs.~\cite{mrsstt,z4f}, the lower 
bound can be relaxed by matching renormalization conditions between a coarse and a fine
lattice in the spirit of ref.~\cite{alpha}).

This has not always been the case at current values of the lattice coupling.
In off-shell correlation functions, at high enough momenta, the dominant 
contributions are perturbative. For example, the leading non perturbative 
contribution of the pseudoscalar correlation function can be easily obtained 
from the Operator Product Expansion (OPE) of the quark propagator, which
has been worked out in refs.~\cite{ppd}. Using either the lattice \cite{grtv} or
the Sum Rules \cite{narison} determination of the chiral condensate and quark mass
values typical of present-day simulations, we estimate that the
non perturbative contribution to the pseudoscalar correlation function 
at $\mu\simeq 2$~GeV is at most 2\%. Therefore, the existence of a  ``renormalization window' 
has been checked in \cite{ggrt1,alphaZ} by 
comparing the $p^2$ dependence of the Green's functions 
to the logarithmic one predicted by continuum perturbation theory
\footnote{Note that continuum perturbation theory converges much better than the
lattice one. Moreover, higher orders have been calculated in the former case, whereas only 1-loop
results are available in lattice perturbation theory.}.
In this sense, at $\beta = 6.0,6.2$ and $6.4$, the scalar density
renormalization constant $Z_S$ is well behaved in a large $\mu$-range,
whereas there is no clear evidence of such a satisfactory renormalization 
window for the RI/MOM renormalization constant of the pseudoscalar
density $Z_P$ (see ref.~\cite{ggrt1} for details).

Recently an alternative interpretation of the observed 
discrepancy between the  RI/MOM and perturbative $Z_P$ results 
has been proposed in ref.~\cite{cyp}. It is suggested that the 
$p^2$ dependence of the pseudoscalar correlation function is 
due to large contributions of the spontaneous chiral symmetry breaking 
terms, which persist at scales used in present day simulations (i.e. $\mu \sim a^{-1}$).
In other words, the lower bound of the renormalization 
window is not adequately satisfied. In order to test this proposal,
the authors of ref.~\cite{cyp} analyze the pseudoscalar correlation 
function, taking into account the leading OPE non-perturbative contribution
and ignoring discretization effects. They find that ``a stricking and unexpected
feature of the lattice data is the very large size of the Goldstone boson
contribution to the pseudoscalar vertex''.

In the present work we critically examine the claim of ref.~\cite{cyp}
by comparing the RI/MOM determination of the ratio $Z_P/Z_S$ to those
obtained from two WIs at $\beta=6.2,6.4$ (the quenched approximation and 
Wilson fermions are implied throughout). WI results are unaffected by 
Goldstone pole contamination but are subject to 
discretization effects. Thus any discrepancy between two WI results gives an estimate of 
$\calo(a)$ effects, whereas an eventually big discrepancy between the WI results and the RI/MOM ones
would signal a Goldstone pole contamination in $Z_P$. We find that individual WI results for
$Z_P/Z_S$
are characterized by errors of about 5\% but they are discrepant to each other at the level of 10-15\%
due to $\calo(a)$ effects. The RI/MOM result obtained at $\mu \sim 2$~GeV is compatible
with the WI ones but has a larger error of about 10-15\%. Thus, contrary to the claim of 
ref.~\cite{cyp}, we conclude that the RI/MOM determination of $Z_P/Z_S$
(and subsequently of $Z_P$) is affected by large discretization errors while any contamination 
from Goldstone pole contributions cannot be discerned. Analogous conclusions, based on an 
analysis of the form factors of the improved quark propagator, have been drawn 
in ref.~\cite{bglm}.

Finally we wish to stress that the WI proposed in this paper (based on off-shell
correlation functions and non-degenerate quark masses) has given a very stable estimate
of $Z_P/Z_S$. In view of this, we prefer to evaluate $Z_P$ from the WI result of
$Z_P/Z_S$ and the RI/MOM value of $Z_S$.

\section{The RI/MOM renormalization scheme and WIs}
\label{sec:basics}
In this section we review the basic characteristics of the RI/MOM
renormalization scheme, implemented with the lattice regularization
\cite{np}. After giving the essential definitions, we discuss the
compatibility of the RI/MOM scheme with (lattice) WIs at large renormalization
scales $\mu$, where the presence of the Goldstone pole becomes negligible.

We start with basic definitions. Given the quark propagator
$\cals (x_1 - x_2;m) =  \langle \psi(x_1) \bar \psi(x_2) \rangle$ and its
Fourier transform
$\cals (p;m) = \int dx \exp(-i p x) \cals (x,m)$ (here $m$ is the quark mass),
we define two ``projections'' (i.e. traces in spin and color space) as   
\bea   
&& \Gamma_\Sigma(p;m) = \frac{-i}{48}   
\Tr \left[ \gamma_\mu \frac{\partial \cals^{-1} (p;m)}{\partial p_\mu} \right]
\nonumber \\   
&& \Gamma_m(p;m) = \frac{1}{12}   
\Tr \left[ \frac{\partial \cals^{-1} (p;m)}{\partial m} \right]   
\eea
We also consider bilinear quark operators of the form
$O_\Gamma(x) = \bar \psi_1(x) \Gamma  \psi_2(x)$ where $\psi_f (x)$ is the
quark field and $\Gamma$ a generic Dirac matrix. For definitiveness we work with
two different flavors $f=1,2$ with corresponding masses $m_1$ and $m_2$.
Specific non-singlet bilinear operators will be denoted as $S(x)$, $P(x)$
(scalar and pseudoscalar densities) and $V_\mu(x)$, $A_\mu(x)$ (vector and
axial currents). Given the insertion of the operator $O_\Gamma(x=0)$ in the
$2$-point fermionic Green's function
\be
G_\Gamma(p) = \int dx_1 dx_2 \exp[-ip (x_1-x_2)]
\langle \psi(x_1) O_\Gamma(0) \bar \psi(x_2) \rangle
\label{eq:gp}   
\ee
and the corresponding amputated correlation function
$\Lambda_\Gamma(p) = \cals ^{-1}(p) G_\Gamma(p) \cals ^{-1}(p)$,
the projected amputated Green's function $\Gamma_\Gamma(p)$ is defined as   
\be   
\Gamma_\Gamma(p) = \frac{1}{12} \Tr \left[\Pj_\Gamma \Lambda_\Gamma(p)\right]   
\label{eq:proj_GF}   
\ee   
where the trace is over spin and color indices and $\Pj_\Gamma$ is the Dirac matrix
which renders the tree-level value of $\Gamma_\Gamma(p)$ equal to unity (i.e.   
it projects out the nominal Dirac structure of the Green function   
$\Lambda_\Gamma(p)$):   
\bea   
&& \Pj_S = I \qquad ; \qquad   
\Pj_P = \gamma_5 \nonumber \\   
&& \Pj_V = \frac{1}{4} \gamma_\mu \qquad ; \qquad   
\Pj_A = \frac{1}{4} \gamma_5 \gamma_\mu   
\label{eq:proj}   
\eea   
Everything defined so far is a bare quantity, assumed to be regularized
on the lattice
\footnote{This means that the integrals of eq.~(\ref{eq:gp}) are really sums   
($a^8 \sum_{x_1,x_2}$) which run over all lattice sites, labelled by $x_1$,   
$x_2$, etc. Also note that the $\mu$-dependence of renormalized correlation functions
is sometimes suppressed; i.e. we use $\hat \Gamma (p;\hat m)$ instead of $\hat \Gamma
(p^2/\mu^2;\hat m^2/\mu^2)$.}
We opt for the lattice regularization scheme with Wilson fermions.
Then the corresponding renormalized quantities in a given mass independent
renormalization scheme are formally given by
\bea   
\hat m_f(\mu) &=& \lim_{a \rightarrow 0} [ Z_m(a\mu) m_f(a) ]
= \lim_{a \rightarrow 0}   
\left[ Z_m \left( m_{0f} - m_C \right) \right]   
\nonumber \\
\hat \cals (p; \hat m_f, \mu) &=& \lim_{a \rightarrow 0}
[Z_\psi (a\mu) \cals (p;m_f,a)]
\nonumber \\
\hat \Gamma_\Sigma (p^2/\mu^2;m_f^2/\mu^2) &=& \lim_{a \rightarrow 0}   
\left[ Z_\psi^{-1}(a\mu) \Gamma_\Sigma (ap,am_f) \right]   
\nonumber \\
\hat \Gamma_m (p^2/\mu^2;\hat m_f^2/\mu^2) &=& \lim_{a \rightarrow 0}   
 \left[ Z_\psi^{-1}(a\mu) Z_m^{-1}(a\mu) \Gamma_m (ap;am_f) \right]   
\nonumber \\
\hat \Gamma_\Gamma(p^2/\mu^2;\hat m_f^2/\mu^2)&=& \lim_{a \rightarrow 0}   
\left[ Z_\psi^{-1}(a\mu) Z_\Gamma(a\mu) \Gamma_\Gamma (ap;am_f) \right]   
\label{eq:form_ren}
\eea
Note that $m_{0f}$ denotes the bare quark mass of a given flavor, 
$m_f = m_{0f} - m_C$ is power subtracted and logarithmically divergent
and $\hat m_f$ is renormalized. Moreover, $Z_\psi^{1/2}$ is the quark
field renormalization and $Z_\Gamma$ the operator renormalization. The functional
dependence of the above expressions is determined by dimensional arguments
and Lorenz invariance.

The renormalization constants in the RI/MOM scheme are defined by
imposing the following off-shell conditions in the deep Euclidean region:
the wave function renormalization $Z_\psi$ is obtained from
\be
\hat \Gamma_\Sigma(p^2/\mu^2;\hat m^2/p^2)   
\Bigg \vert_{\begin{array}{c}   
p^2 = \mu^2 \\   
\hat m^2 = 0 \end{array}}   
=  \lim_{a \rightarrow 0} \lim_{m \rightarrow 0}   
Z_\psi^{-1}(a\mu) \Gamma_\Sigma(a\mu; a^2 m^2)   
=1 
\label{eq:rcpsi}   
\ee   
the quark mass renormalization $Z_m$ is achieved through
\be
\hat \Gamma_m(p^2/\mu^2;\hat m^2/p^2)   
\Bigg \vert_{\begin{array}{c}   
p^2 = \mu^2 \\   
\hat m^2 = 0 \end{array}}   
= \lim_{a \rightarrow 0} \lim_{m \rightarrow 0} Z_\psi^{-1}(a\mu)   
Z_m^{-1}(a\mu) \Gamma_m(a\mu; a^2 m^2)   
=1   
\label{eq:rcmass}   
\ee
and for the bilinear operator $Z_\Gamma$ is obtained by imposing the renormalization
condition   
\be   
\hat \Gamma_\Gamma (p^2/\mu^2;\hat m^2/p^2)   
\Bigg \vert_{\begin{array}{c}   
p^2 = \mu^2 \\   
\hat m^2 = 0 \end{array}}   
= \lim_{a \rightarrow 0} \lim_{m \rightarrow 0}   
Z_\psi^{-1}(a\mu) Z_\Gamma(a\mu) \Gamma_\Gamma (a\mu; a^2 m^2) = 1   
\label{eq:rcop}   
\ee   
For any fixed scale $\mu$, this procedure removes the UV divergences from all
Green functions and thus renormalization is achieved.
However, renormalization conditions must also be chosen so that the resulting
renormalized operators transform ``correctly'' under the chiral group; i.e.
they should belong to the same chiral representation as the nominal bare operators.
This additional requirement, which ensures that chiral symmetry survives
renormalization, is only true at large enough scales $\mu$. At small scales
symmetry violating effects due to spontaneous chiral symmetry   
breaking in QCD should appear as $\Lambda_{QCD}$-dependent form factors   
in the renormalized Green functions   
$\hat \Gamma_\Gamma (p^2/\mu^2;\hat m^2/p^2;\Lambda_{QCD}/p^2)$.   
Moreover, in practical simulations, usually performed at (light) non-zero quark
mass, explicit chiral symmetry breaking form factors, proportional to   
$\hat m(\mu)$ will also be present. Both types of form factors become   
negligible if the scale $\mu$ is adequately large. Therefore, the requirement
$\Lambda_{QCD} \ll \mu$ has to be satisfied in principle, so that
quantities renormalized in the RI/MOM scheme respect chiral symmetry.
Moreover, since one is usually working at fixed UV cutoff in simulations,
the requirement $\mu \ll \calo(a^{-1})$ must be satisfied in order to control
discretization errors. Thus, the reliability of the RI/MOM scheme on the
lattice is ensured provided one is working in a renormalization window
$\Lambda_{QCD} \ll \mu \ll \calo(a^{-1})$.

The above general statements have been shown to be true in ref.~\cite{np}   
(and generalized for some cases of additive renormalization of dimension-six   
four-fermion operators in refs.~\cite{testa,z4f}).   
In particular it has been shown that the RI/MOM scheme is always   
compatible with vector WIs, whereas it is only compatible with the axial   
WIs at large scales $\mu$. The key observation is that the vector WIs
\bea
\hat \Gamma_S \left( p; \hat m \right) 
&=& \hat \Gamma_m \left( p; \hat m \right)
\nonumber \\
\hat \Gamma_V \left( p; \hat m \right)   
&=& \hat \Gamma_\Sigma \left( p; \hat m \right) 
\label{eq:vwi-rimom}
\eea
at momentum $p^2 = \mu^2$, are automatically satisfied by the renormalized
quantities $\hat \Gamma_\Sigma$, $\hat \Gamma_m$, $\hat \Gamma_V$ and
$\hat \Gamma_S$ determined in the RI/MOM scheme by eqs.~(\ref{eq:rcpsi}),
(\ref{eq:rcmass}) and (\ref{eq:rcop}). Conversely, if we use the RI/MOM
scheme to fix, say, $\hat \Gamma_\Sigma$ and $\hat \Gamma_m$
(i.e. $Z_\psi$ and $Z_m$), then $Z_V$ and $Z_S$ (through the identity
$Z_S = Z_m^{-1}$), as
fixed by the WIs~(\ref{eq:vwi-rimom}), are compatible to those obtained from
the RI/MOM condition~(\ref{eq:rcop}).

The compatibility of axial WIs to RI/MOM is more intricate;   
the proof would proceed in exactly the same way as in the vector case,   
if it were not for the extra terms on the l.h.s. of WIs
\bea
\hat \Gamma_P \left( p; \hat m \right) + \hat m   
\dfrac{\partial \hat \Gamma_P \left( p; \hat m \right)}{\partial \hat m}   
&=& \hat \Gamma_m \left( p; \hat m \right)   
\nonumber \\
\hat \Gamma_A \left( p; \hat m=0 \right) + \lim_{q_\rho\rightarrow 0} \frac{q_\mu}{48}   
\Tr \left [ \gamma_5 \gamma_\rho \dfrac{\partial   
\hat \Lambda_A^\mu (p+q/2,p-q/2; \hat m=0)}{\partial q_\rho} \right]   
&=&  \hat \Gamma_\Sigma \left( p; \hat m=0 \right)  
\label{eq:awi-rimom}
\eea
which is the axial analogue to the vector WIs~(\ref{eq:vwi-rimom}). Note that 
the second term of the l.h.s. of WIs~(\ref{eq:awi-rimom}) does not vanish in
the chiral limit, due to the presence of a Goldstone pole. As shown in
ref.~\cite{np} these extra terms are negligible in the deep Euclidean region
$\Lambda_{QCD} \ll p$. In this limit, the WIs and the RI/MOM conditions are
compatible, just like the vector case. This limit corresponds to the
lower bound of the renormalization window requirement, inherent in the
RI/MOM scheme. Clearly, all statements made thus far are true up to
discretization effects, which are present both in the RI/MOM and the WI
implementations on the lattice. We will address this problem at a later stage
of this work.
What we need to consider at present, is that RI/MOM results are reliable
in the renormalization window $\Lambda_{QCD} \ll p \ll \calo(a^{-1})$,
whereas WI results are reliable in the window $p \ll \calo(a^{-1})$.

The main aim of the present work is to establish whether
there is significant Goldstone pole contamination of the lattice RI/MOM
results obtained thus far. In particular, the RI/MOM determination of 
$Z_P/Z_S$ at current $\beta$ values and momenta does not display a clear
plateau; see refs.~\cite{np}-\cite{sfqcd}. The problem may be either due to
the Goldstone pole problem or, as claimed in  ref.~\cite{bglm}, due to discretization
effects\footnote{Moreover, the 1-loop Perturbation Theory PT (and Boosted PT -BPT)
result for $Z_P$ is in stark disagreement with the RI/MOM estimate. This comparison,
however, does not reveal much, as the BPT result also suffers from $\calo(g_0^4)$
errors.}.
Since the WI determination of the ratio $Z_P/Z_S$ does not suffer from
Goldstone pole contamination, comparison of the WI and the RI/MOM results
for this quantity should offer a way of estimating this effect. For example,
the WI result from eqs.~(\ref{eq:vwi-rimom}) and  (\ref{eq:awi-rimom}) is
\be
\dfrac{Z_P}{Z_S} = \dfrac{ \dfrac{\Gamma_S}{\Gamma_P} }
{1 + \dfrac{m}{\Gamma_P}\dfrac{\partial \Gamma_P}{\partial m} }
\label{wi:teo}
\ee
whereas the RI/MOM result from eq.~(\ref{eq:rcop}) is
\be
\label{eq:momteo}
\dfrac{Z_P}{Z_S} = \dfrac{\Gamma_S}{\Gamma_P}
\ee
The denominator on the r.h.s. of the WI~(\ref{wi:teo}) is missed by the
RI/MOM eq.~(\ref{eq:momteo}) and this results to a Goldstone pole
contamination of the latter.
This particular WI determination, however, is difficult to implement in
practice (e.g. realization of derivatives w.r.t. the quark mass in lattice
numerical simulations etc.). In the next section we discuss several WIs, which
are equivalent (up to discretization effects) but more useful in practice.

\section{The ratio $Z_P/Z_S$ from WIs}
\label{subsec:zpzs}
We will now review three methods, based on WIs, for the determination of the   
scale independent ratio $Z_P/Z_S$. A first method consists in computing
$Z_P/Z_S$ as the ratio of the PCAC current quark mass to the bare (subtracted)
quark mass \cite{alpha}.
In standard notation (see refs.~\cite{boc}, \cite{clv}) the
current quark mass for Wilson fermions is obtained from the standard PCAC
relation (an axial WI, valid $\forall x \ne 0$):
\be
2 \left[m_0 - \overline m(m_0) \right] =
Z_A \dfrac{\partial_0 \,\, \int d^3x \langle A_0(x) P(0) \rangle}   
{\int d^3 x \langle P(x) P(0) \rangle}
\label{eq:2rho}   
\ee
whereas the mass power subtraction $m_C$ can be obtained by linearly
extrapolating either the square of the pion mass $m_\pi^2$ or
$\left[m_0 - \overline m(m_0) \right]$ to their vanishing value. 
Then vector
and axial WIs (see refs.~\cite{ks,boc,clv}) determine the quark mass
renormalization
\be
\hat m = Z_S^{-1} \left[ m_0 - m_C \right]
= Z_P^{-1} \left[ m_0 - \overline m \right]
\label{eq:qmassren}
\ee
and thus by computing $ \left[ m_0 - \overline m \right]$ at several $m_0$'s
the ratio $Z_P/Z_S$ is obtained as the slope of
\be
m_0 - \overline m = \dfrac{Z_P}{Z_S}  \left[ m_0 - m_C \right]\; .
\label{eq:zpzsslope}
\ee
This determination does not depend on the critical quark mass $m_C$. It does,
however, depend on the determination of $Z_A$. Since the methods relies
on the WI (\ref{eq:2rho}) on hadronic states, we will label it WIh.

A second method consists in combining eq.~(\ref{eq:zpzsslope})
with the axial WI
\bea
\label{eq:aw-meth2}
(m_{01} + m_{02} - 2 \overline m) \Gamma_P\left( ap; am_1, am_2 \right) &=&
  \dfrac{1}{12} \Tr \cals^{-1} \left (ap; am_1 \right)   
+ \dfrac{1}{12} \Tr \cals^{-1} \left (ap; am_2 \right)
\eea
(considered in the mass-degenerate case), in order to obtain
\be
\dfrac{Z_P}{Z_S} 
= \dfrac{\Tr \cals^{-1}(ap;am)}
{12 \left [ m_0 - m_C \right ]  \Gamma_P(ap; a m, a m) }
\label{eq:zpzswihat}
\ee 
This determination requires the projected amputated correlation functions
and the quark propagators in momentum space. We anticipate at this stage, that
this method fails in practice. This is due to the large discretization errors
of  the quark propagator (a detailed investigation of these effects, which
remain large even when Clover improvement is implemented beyond tree-level,
has been performed in ref.~\cite{bglm}).

A third method consists in combining eq.~(\ref{eq:aw-meth2})
for the degenerate mass case with the vector WI
\bea
(m_{02} - m_{01}) \Gamma_S \left(ap; am_1, am_2 \right) &=&   
- \dfrac{1}{12} \Tr \cals^{-1} \left (ap; am_1 \right)   
+ \dfrac{1}{12} \Tr \cals^{-1} \left (ap; am_2 \right)        
\label{eq:vw-meth2}
\eea   
so as to eliminate
the quark propagators. Using also eq.~(\ref{eq:zpzsslope}) we obtain
\be
\dfrac{Z_P}{Z_S} = \dfrac{\left( m_1 - m_2 \right)
\Gamma_S \left(ap;am_1,am_2\right)}
{m_1 \Gamma_P\left(ap;am_1,am_1\right) - m_2\Gamma_P\left(ap;am_2,am_2\right)}
\label{eq:zpzswigood}
\ee
The advantage of this determination is that the large ${\cal O}(ap)$
discretization errors,
present in the quark propagator (cf. ref.~\cite{bglm}), are
eliminated. Since this determination depends on correlation functions of
external quark states, we will label it WIq. Note that this method is
characterized by non-degenerate quark masses.

Finally, we mention that a WI with hadronic states has also been used for the
computation of the ratio $Z_P/Z_S$; see refs.~\cite{clv,mpsv}.

\section{Results}
\label{sec:results}
Our results are based on earlier simulations; all technical details can
be found in refs.~\cite{ggrt1,ggrt2}. Here we just mention that the dataset
we analyze has been obtained with the Wilson (unimproved) action, at couplings
$\beta = 6.2$ and $6.4$ and at several hopping parameters in the range of the
strange quark mass. All results shown here are extrapolated to the chiral
limit. The RI/MOM results are those of ref.~\cite{ggrt1} (which agree with
the ones of ref.~\cite{sfqcd}) whereas the WIh and WIq results are new.
We mostly show the Wilson $\beta = 6.2$ case
\footnote{Note that our $\beta = 6.0$ dataset has also been analyzed, with
qualitatively similar results.}.

\begin{figure}[htb]
\begin{center}
\caption{\it{The value of $2[m_0 - \overline m]/Z_A$, obtained from
${\rm eq.~(\ref{eq:2rho})}$, as a function of the Wilson hopping
parameter $1/K$ at $\beta = 6.2$. The slope of the linear fit
is the ${\rm WIh}$ result for $Z_P/[Z_S Z_A]$, according to
${\rm eq.~(\ref{eq:zpzsslope})}$. The $K_C$ value implicit in the abscissa axis
is determined from the vanishing of $m_\pi^2$ (linear in $1/K$).}}
\label{fig:2rho}
\vskip 0.3cm
\ifigsm{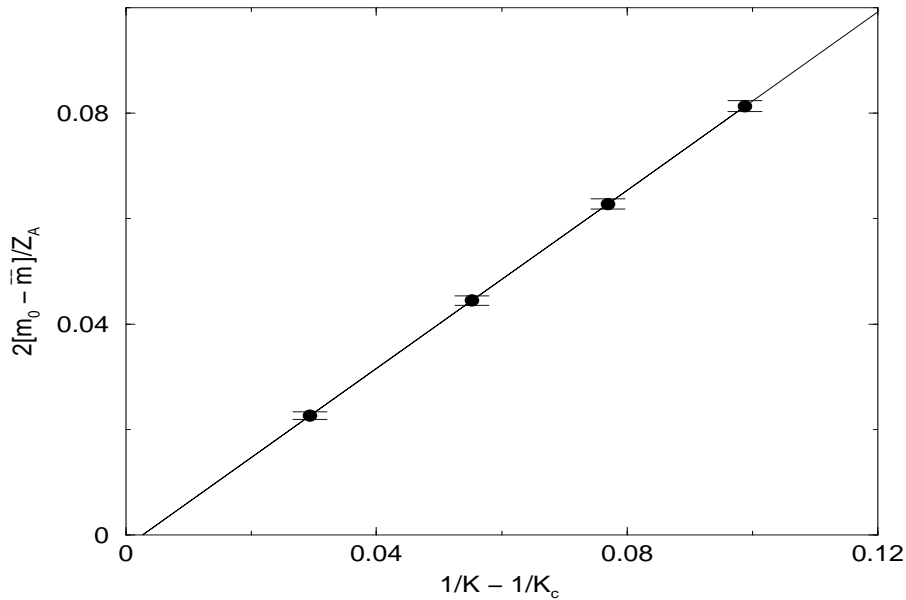}
\end{center}
\end{figure}

In figure \ref{fig:2rho} we show the data and linear
fit characteristic of the method WIh. The slope of the fitting line
is $Z_P/[Z_S Z_A]$. In order to extract $Z_P/Z_S$ from the slope,
we use the RI/MOM estimate for $Z_A$,
as given in ref.~\cite{ggrt1}. This is in accordance (with larger errors) to
the RI/MOM result of ref.~\cite{sfqcd}.

In Fig.~\ref{fig:wi_klm}(a) we show our $\beta = 6.2$ results obtained
from WI~(\ref{eq:zpzswihat}), as a function of the (lattice) scale
\be
a^2 {\overline \mu}^2 = \sum_\nu \sin^2(a\mu_\nu)
\ee
Due to enormous systematic errors, our results
range over a large interval and no plateau is seen. The tree-level value of
eq.~(\ref{eq:zpzswihat}) is given by
\be
\dfrac{Z_P}{Z_S} = 1 + \dfrac{2 \sum_\nu \sin^2 \left(a p_\nu /2 \right)}
{a m }
\label{eq:wi_klm}
\ee
where the second term on the r.h.s. is the tree-level ${\cal O}(a)$ effect
arising from the propagator sum in the numerator of eq.~(\ref{eq:zpzswihat}).
In Fig.~\ref{fig:wi_klm}(b) we have corrected our results by this factor,
in the spirit of KLM tree-level improvement
(cf. ref.~{\cite{klm}). Although this factor accounts for a big
part of the discretization errors, the remaining effects are still large and
render the method unreliable.

\begin{figure}[htb]
\begin{center}
\caption{\it{$Z_P/Z_S$ results at $\beta = 6.2$.
${\rm (a)}$ Results obtained from ${\rm WI~(\ref{eq:zpzswihat})}$;
${\rm (b)}$ Results corrected by the ${\rm KLM}$ factor of
${\rm eq.~(\ref{eq:wi_klm})}$.}}
\label{fig:wi_klm}
\ifig{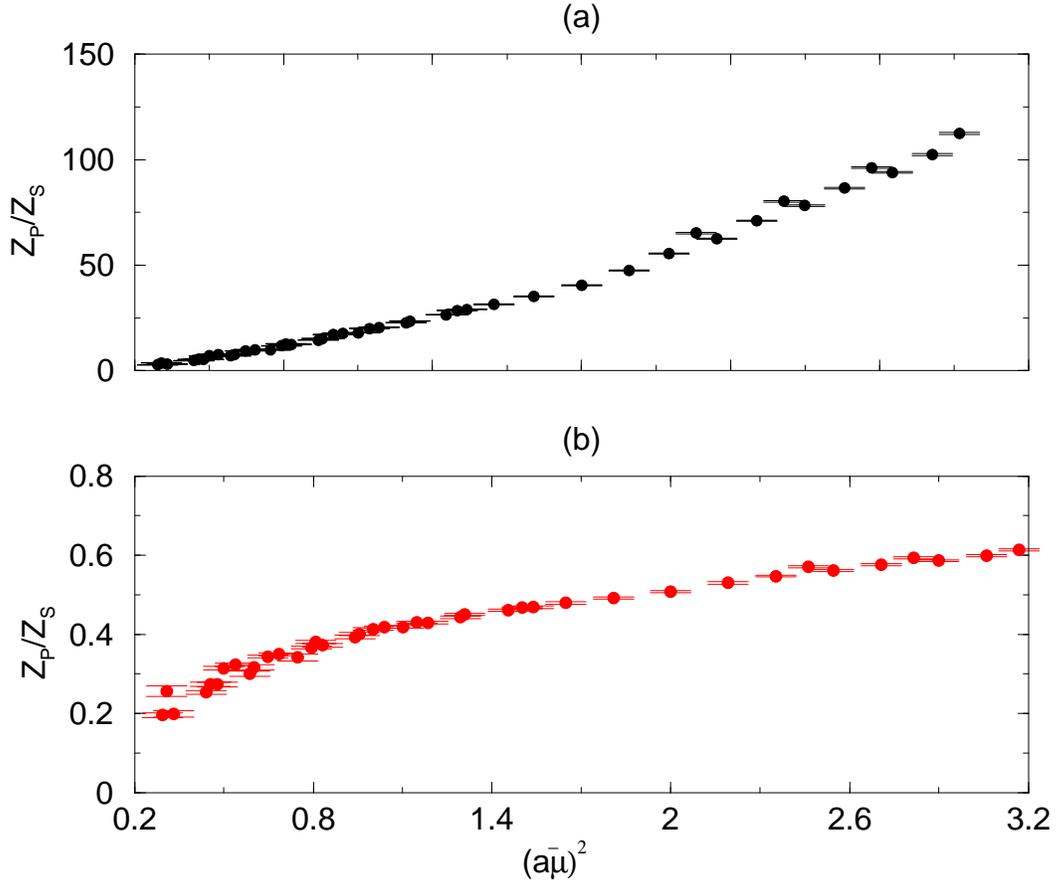}
\end{center}
\end{figure} 
In Fig.~\ref{fig:Zp_Zs} we compare the $\beta = 6.2$ results for the
RI/MOM and WIq determinations, as a function of the (lattice) scale
$a^2 {\overline \mu}^2$.
It is clear that at small scales there is a large discrepancy due to the
sizeable Goldstone pole contamination of the RI/MOM method, which decreases
with increasing scale. At large scales it is impossible to discern whether the
discrepancy is due to a remnant Goldstone pole contamination or the
discretization effects. Moreover, even at larger scales, the renormalization
window of the RI/MOM method, compared to the plateau displayed by the WIq
results, is rather poor. In ref.~\cite{ggrt1}, the central values of 
the RI/MOM results were taken at $(a {\overline \mu})^2\simeq 0.8$
and the systematic errors estimated as the spread of the 
values in the region $0.8\leq (a {\overline \mu})^2\leq 1.5$. 
This choice takes into account the instability of the renormalization window,
yielding a RI/MOM result with large errors. On the contrary, the WI data
displays a good plateau already at rather small scales. Our choice
$(a {\overline \mu})^2 \in [0.2,0.8]$ for the WIq plateau gives very accurate
results.

\begin{figure}[htb]
\begin{center}
\caption{\it{$Z_P/Z_S$ results at $\beta = 6.2$.
${\rm (a)}$ ${\rm WIq}$ and ${\rm RI/MOM}$ determinations as a function of the
renormalization scale. ${\rm (b)}$ Enlargement of ${\rm (a)}$, highlighting
the $\mu$ values where scale independence is expected.}}
\label{fig:Zp_Zs}
\ifig{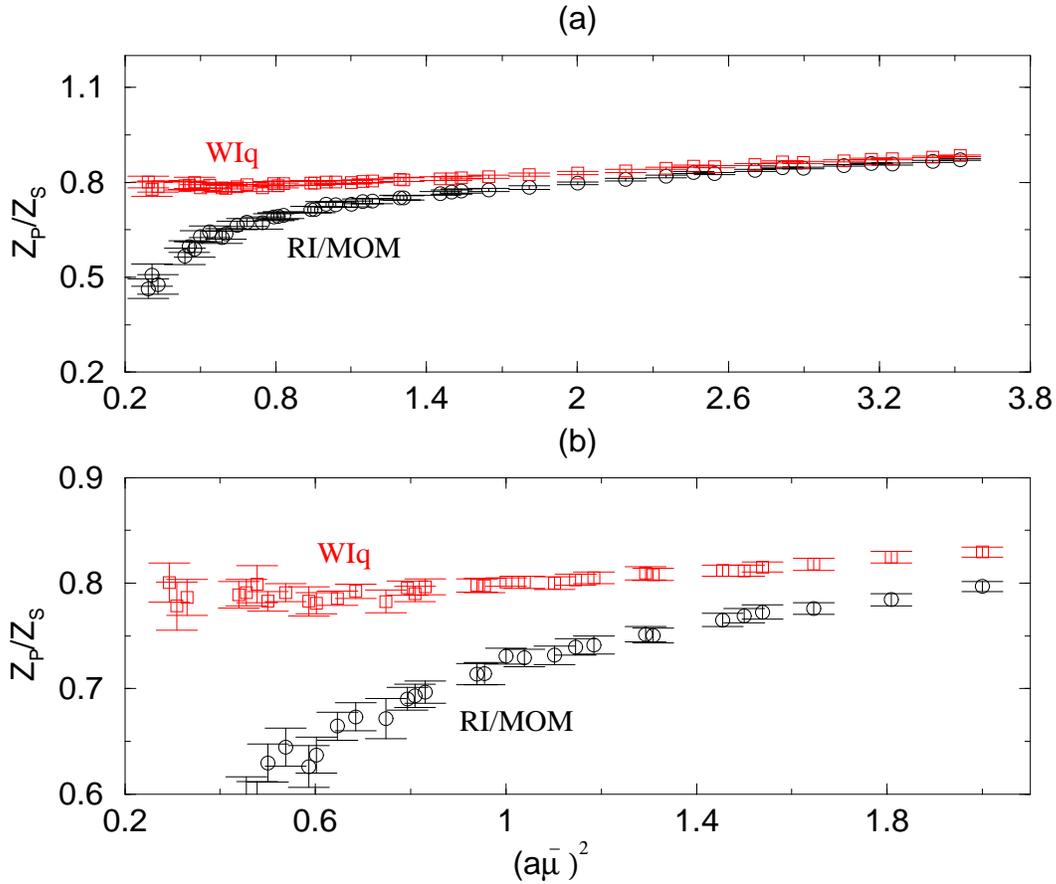}
\end{center}
\end{figure}
In order to estimate the discretization errors, we collect in
Fig.~\ref{fig:Zp_Zs_altri_6264}, the various determinations of $Z_P/Z_S$ at
$\beta = 6.2$ and $6.4$. The most accurate estimates are those based on the two
WIs (WIh and WIq). They are not, however fully compatible, due to
discretization effects (recall that WI results do not suffer from the Goldstone
pole problem). This indicates that these effects introduce a
systematic error of $\simeq 15\%$ at $\beta=6.2$, which decreases to
$\simeq 10\%$ at $\beta = 6.4$. The RI/MOM determination, which is far less accurate
due to the instability of the renormalization window, is nevertheless in the
same range of values. Thus, it is clear that any remnant Goldstone pole
contamination in the RI/MOM result is not superior in magnitude to the uncertainties
arising from discretization effects. Also, the fact that the spread of the
various determinations of $Z_P/Z_S$ decrease with increasing $\beta$,
indicates that the discrepancies are due to (decreasing) finite cutoff effects.
\begin{figure}[htb]
\begin{center}
\caption{\it{$Z_P/Z_S$ results ${\rm (a)}$ at $\beta = 6.2$;
${\rm (b)}$ at $\beta=6.4$.
Comparison of non-perturbative ${\rm (WIq,WIh,RI/MOM)}$ and perturbative
${\rm (BPT,PT)}$ determinations.}}
\label{fig:Zp_Zs_altri_6264}
\ifigsm{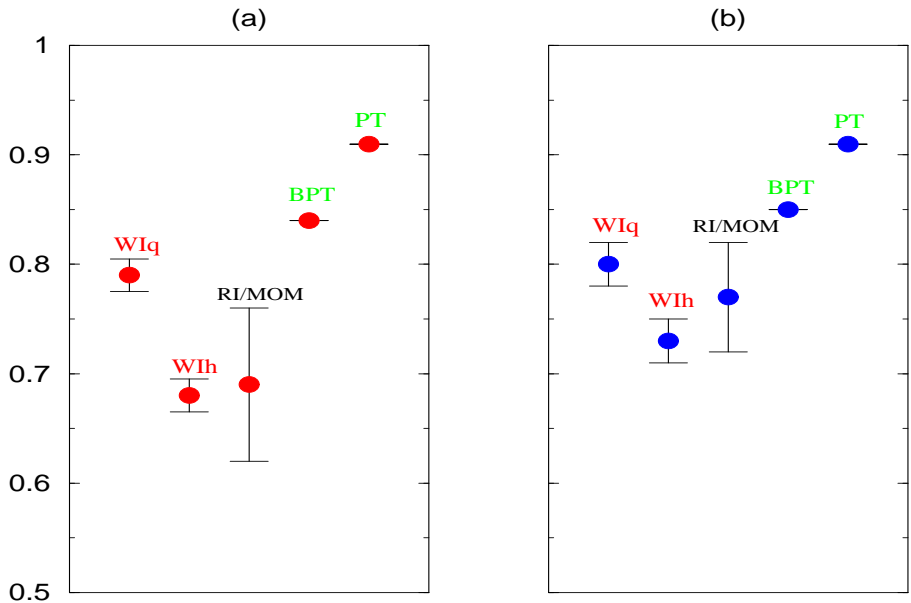}
\end{center}
\end{figure}
\begin{table}
\centering
\begin{tabular}{||c|c|ccc||}
\hline\hline
$\beta$ &
Method & $Z_P$ & $Z_S$ & $Z_P/Z_S$
\\
\hline\hline
      & PT                   & 0.78    & 0.86    & 0.91    \\ 
 6.2  & BPT                  & 0.65    & 0.77    & 0.84    \\ 
      & RI/MOM \cite{ggrt1}   & 0.50(5) & 0.72(1) & 0.69(7) \\ 
      & WIq [This work]      & 0.57(1) &   -     & 0.79(2) \\
      & WIh [This work]      & 0.50(2) &   -     & 0.68(2) \\ 
\hline
      & PT                   & 0.79    & 0.86    & 0.91   \\ 
 6.4  & BPT                  & 0.67    & 0.79    & 0.85   \\ 
      & RI/MOM \cite{ggrt1}  & 0.57(4) & 0.74(1) &  0.77(5)\\  
      & WIq [This work]      & 0.60(1) &   -     & 0.80(2) \\ 
      & WIh [This work]      & 0.54(2) &   -     & 0.73(2) \\ 
\hline\hline
\end{tabular}
\caption{\it Perturbative (PT and BPT) and non-perturbative (RI/MOM, WIq, WIh)
estimates of renormalization constants. The RI/MOM values are at $\bar \mu^2a^2=0.8$.
The WIq and WIh results are obtained from eq.~(\ref{eq:zpwi}).}
\label{tab:Z(m=0)}
\end{table}

Our new WI results demonstrate that the discrepancy of $10-15\%$ for $Z_P/Z_S$ carries over to $Z_P$,
computed (non-perturbatively) from 
\be
Z_P = \left[ \dfrac{Z_P}{Z_S} \right]^{{\rm WI}}Z_S^{{\rm RI/MOM}}\; .
\label{eq:zpwi}
\ee
In practice we use the WIq determination for the ratio $Z_P/Z_S$ in the above equation; i.e.
combining eq.~(\ref{eq:zpzswigood}) with the RI/MOM renormalization condition (\ref{eq:rcop})
for the scalar density, we have
\be
Z_P = Z_\psi \dfrac{m_1 - m_2 }
{m_1 \Gamma_P\left(a\mu;am_1,am_1\right) - m_2\Gamma_P\left(a\mu;am_2,am_2\right)}\; .
\label{eq:zpgood}
\ee
The wave-function renormalization constant $Z_\psi$ has been computed in ref.~\cite{ggrt1}. 
In Table \ref{tab:Z(m=0)} we collect our results for $Z_P$, $Z_S$ and their ratio.
Consequently, at these $\beta$ values, determinations of the quark mass (and chiral condensate)
which differ in the choice of non-perturbative renormalization (i.e. WIq, WIh, RI/MOM),
would also display the same variations
\footnote{In refs.~\cite{ggrt2,grtv}, a good agreement was observed
between the quark mass (and chiral condensate), renormalized with $Z_S$ and that
renormalized with $Z_P$ in the RI/MOM scheme.
This is equivalent to the good agreement between the central values of the
WIh and RI/MOM estimates of $Z_P/Z_S$. It is clear from fig.~\ref{fig:Zp_Zs_altri_6264} that this
is accidental.}.
These differences should vanish upon extrapolation to the continuum limit.

It is interesting to note that, upon expressing the first of WIs~(\ref{eq:vw-meth2}) in terms
of renormalized quantities and using the RI/MOM renormalization condition for $\Gamma_S$, the relation
\be
\left[\hat m_1(\mu) - \hat m_2(\mu)\right]^{\RIMOM} = 
\left[ \dfrac{1}{12} \Tr \hat \cals^{-1} \left (\mu; \hat m_1 \right)   
- \dfrac{1}{12} \Tr \hat \cals^{-1} \left (\mu; \hat m_2 \right) \right]^{\RIMOM}
\label{eq:massnew}
\ee
can readily be obtained. This gives a non-perturbative
quark-mass determination which does not require knowledge of $Z_S$ (or $Z_P$);
cf. ref.~\cite{bglm}. Moreover, it does not suffer from large $\calo(ap)$ discretization effects
(they cancel in the difference of quark propagators at any quark mass values), which would
otherwise have to be isolated by chiral extrapolation. 
This determiantion does however require the computation of the wave function renormalization
$Z_\psi$ in the RI/MOM scheme. Since this is a mass-independent renormalization scheme,
the above can only be implemented in the limit $m^2/\mu^2\ll 1$.

\section*{Acknowledgments}
We wish to thank V.~Lubicz, G.~Martinelli and S.~Sint for numerous illuminating discussions.
A.V. thanks the DESY Theory Group for its hospitality during the early
stages of this work. L. G. has been supported in part under DOE grant DE-FG02-91ER40676.


\begin{thebibliography}{999}      
\bibitem{np}   
G.~Martinelli et al., Nucl. Phys. B445 (1995) 81.    
\bibitem{ggrt1}
V.~Gimenez, L.~Giusti, F.~Rapuano and M.~Talevi,
Nucl. Phys. B531 (1998) 429.
\bibitem{sfqcd}
M.~G\"ockeler at al., Nucl. Phys. B544 (1999) 699.
\bibitem{z4f}
A.~Donini et al., Eur. Phys. J. C10 (1999) 121.
\bibitem{mrsstt}
G.~Martinelli et al., Phys. Lett. B411 (1997) 141.
\bibitem{alpha}
K.~Jansen et al., Phys. Lett. B372 (1996) 275;\\
M.~L\"uscher et al., Nucl.Phys. B491 (1997) 323.
\bibitem{ppd}
H.D.~Politzer, Nucl.Phys. B117 (1976) 397;\\
P.~Pasqual and E.~de~Rafael, Z. Phys. C12 (1982) 127;\\
M.J.~Lavelle and M.~Oleszczuk, Phys. Lett. B275 (1992) 133.
\bibitem{grtv}
L.~Giusti, F.~Rapuano, M.~Talevi and A.~Vladikas,
Nucl. Phys. B538 (1999) 249.
\bibitem{narison}
S.~Narison, 
Nucl. Phys. (Proc. Suppl.) 86 (2000)~242 and references therein. 
\bibitem{alphaZ}
S.~Capitani et al., Nucl. Phys. B544 (1999) 669.
\bibitem{cyp}   
J-R.~Cudell, A.~Le~Yaouanc and C.~Pittori,
Phys. Lett. B454 (1999) 105.   
\bibitem{bglm}   
D.~Becirevic, V.~Gimenez, V.~Lubicz and G.~Martinelli, 
Phys. Rev. D61 (2000) 114507.
\bibitem{testa}
M.~Testa, Nucl. Phys. B (Proc. Suppl.) 63 (1998) 38;\\
JHEP 04 (1998) 002.
\bibitem{ggrt2}   
V.~Gimenez, L.~Giusti, F.~Rapuano and M.~Talevi,
Nucl. Phys. B540 (1999) 472.
\bibitem{boc}   
M.~Bochicchio et al., Nucl. Phys. B262 (1985) 331.   
\bibitem{clv}   
M~.Crisafulli, V.~Lubicz and A.~Vladikas, Eur.~Phys.~J. C4 (1998) 145.   
\bibitem{ks}   
L.H.~Karsten and J.~Smit, Nucl. Phys. B183 (1981) 103.   
\bibitem{mpsv}
L.~Maiani and G.~Martinelli, Phys. Lett. B178 (1986) 265;\\
G.Martinelli, S.~Petrarca, C.T.~Sachrajda and A.~Vladikas
Phys. Lett. B311 (1993) 241;\\ Phys. Lett. B317 (1993) 660;\\
D.S.~Henty et al. (UKQCD collaboration), Phys. Rev. D51 (1995) 5323.
\bibitem{klm}
G.P.~Lepage, Nucl. Phys. (Proc. Suppl.) 26 (1992) 45;\\
A.S.~Kronfeld, Nucl. Phys. (Proc. Suppl.) 30 (1993) 445;\\
P.B.~Mackenzie, Nucl. Phys. (Proc. Suppl.) 30 (1993) 35.
\end{thebibliography}
\end{document}